\documentclass[12pt]{article}

\begin{document}

\title{\bf Energy-Momentum Distribution in Weyl Metrics}

\author{M. Sharif \thanks{e-mail: hasharif@yahoo.com} and Tasnim Fatima\\
Department of Mathematics, University of the Punjab,\\
Quaid-e-Azam Campus, Lahore-54590, Pakistan.}

\date{}

\maketitle

\begin{abstract}
In this paper, we evaluate energy and momentum density
distributions for the Weyl metric by using the well-known
prescriptions of Einstein, Landau-Lifshitz, Papaterou and
M$\ddot{o}$ller. The metric under consideration is the static
axisymmetric vacuum solution to the Einstein field equations and
one of the field equations represents the Laplace equation. Curzon
metric is the special case of this spacetime. We find that the
energy density is different for each prescription. However,
momentum turns out to be constant in each case.
\end{abstract}


\section{Introduction}

The concept of energy, momentum and angular momentum has always an
important character in both Classical and Quantum Physics. In
Special Relativity, the density of energy and momentum form a
second rank tensor field $T^b_a$ whose divergence vanishes.
Serious difficulties in connection with its notion arise in the
theory of General Relativity (GR). The energy problem is the
oldest [1] and one of the most difficult problems of classical GR.

The law of conservation of energy in GR, $T^b_{a;b}=0$, can be
written in the form
\begin{equation}
(T^b_a+t^b_a)_{,b}=0,
\end{equation}
where $t^b_a$ is the energy-momentum pseudo tensor of the
gravitational field. Because of the way we defined $t^b_a$, it is
not a tensor - the procedure of picking out a partial derivative
from a covariant one is not invariant. By using a superpotential
$H^{bc}_a,~H^{bc}_{a,c},~t^b_a$ is defined non-covariantly from
the very beginning. In the same way, the canonical energy-momentum
pseudo tensor is not a tensor, because to form it we take a
partial (non-covariant) derivative of the Lagrangian.

There have been many attempts to resolve the energy problem in GR.
Penrose [2] introduced a definition of quasi-local mass in order
to discuss gravitational energy in the framework of GR.  The
definition applies to space-like surfaces of spherical topology
and, through construction intrinsic to the two-surface, produces a
quantity to be termed as the Pensrose mass within the two surface.
In linearised theory this is the same as the norm of the
energy-momentum of the sources within the two-surface. So far most
of the applications have been made in asymptotic regions, null and
space-like infinity for asymptotically flat spaces and time-like
infinity for asymptotically anti-de Sitter spaces [2,3].

Jeffryes [4] applied the construction of quasi-local mass in the
Newtonian limit of GR with perfect fluids using the formulation of
Futamase et al. [5]. Bartnik [6] presented a new defintion of
quasi-local mass. Dirac and Arnowitt et al. [7] succeeded in
putting the exact theory of gravitation in Hamiltonian form.

There have been series of some other attempts [8]-[13] to evaluate
the energy-momentum distribution. In this series, the first
attempt was made by Einstein [8] who suggested an expression for
energy-momentum density. After this, many prescriptions, followed
by Landau-Lifshitz [9], Papapetrou [10], Bergman [11], Tolman [12]
and Weinberg [13], were proposed. The main problem with these
definitions is that they are coordinate dependent. These
prescriptions give meaningful results only when calculations are
performed in Cartesian coordinates.

M\"{o}ller [14,15] proposed an expression which is the best to
make calculations in any coordinate system. He claimed that his
expression would give the same values for the total energy and
momentum as the Einstein's energy-momentum complex for a closed
system. However, M$\ddot{o}$ller's energy-momentum complex was
subjected to some criticism [2], [15]-[17]. Komar's [16]
prescription, though not restricted to the use of Cartesian
coordinates, is not applicable to non-static spacetimes. Thus each
of these energy-momentum complex has its own drawbacks. As a
result, these ideas of the energy-momentum complex were severally
criticized.

Virbhadra et al. [18]-[22] and Xulu [23,24] explored several
examples of the spacetimes and found that different prescriptions
could provide exactly the same energy-momentum distribution.
Virbhadra concluded that Einstein's prescription might provide the
best results among all the known prescriptions for the
energy-momentum distribution of a given spacetime. In a recent
paper, Lessner [25] pointed out that the M$\ddot{o}$ller's
energy-momentum prescription is a powerful concept of energy and
momentum in GR.

In recent papers, Sharif [26-28] has explored some examples of
spacetimes in which energy-momentum density components turn out to
be different for different prescriptions. In this paper, we use
energy-momentum complexes of Einstein, Ladau-Lifshitz, Papapetrou
and M$\ddot{o}$ller to evaluate energy-momentum density components
of the Weyl metric. The paper is formulated as follows. In the
next section, we shall describe the Weyl metric and transform it
into Cartesian coordinates. In sections 3, 4, 5 and 6, we shall
evaluate energy and momentum densities using the prescriptions of
Einstein, Landau-Lifshitz, Papapetrou and M\"{o}ller respectively.
In the last section, we shall discuss the results.

\section{The Weyl Metrics}

Static axisymmetric solutions to the Einstein field equations are
given by the Weyl metric [30]
\begin{equation}
ds^2=e^{2\psi}dt^2-e^{2(\gamma-\psi)}(d\rho^2+dz^2)
-\rho^2e^{-2\psi}d\phi^2
\end{equation}
in the cylindrical coordinates $(\rho,~\phi,~z)$. Here $\psi$ and
$\gamma$ are functions of coordinates $\rho$ and $z$. The metric
functions satisfy the following differential equations
\begin{eqnarray}
\psi_{\rho\rho}+\frac{1}{\rho}\psi_{\rho}+\psi_{zz}=0,\\
\gamma_{\rho}=\rho(\psi^2_{\rho}-\psi^2_{z}),\quad
\gamma_{z}=2\rho\psi_{\rho}\psi_{z}.
\end{eqnarray}
It is obvious that Eq.(3) represents the Laplace equation for
$\psi$. Its general solution, yielding an asymptotically flat
behaviour, will be
\begin{equation}
\psi=\sum^\infty_{n=0}\frac{a_n}{r^{n+1}}P_n(\cos\theta),
\end{equation}
where $r=\sqrt{\rho^2+z^2},~\cos\theta=z/r$ are Weyl spherical
coordinates and $P_n(\cos\theta)$ are Legendre Polynomials. The
coefficients $a_n$ are arbitrary real constants which are called
{\it Weyl moments}. It is mentioned here that if we take
\begin{eqnarray}
\psi=-\frac{m}{r},\quad\gamma=-\frac{m^2\rho^2}{2r^4},\quad
r=\sqrt{\rho^2+z^2}
\end{eqnarray}
then the Weyl metric reduces to special solution of Curzon metric
[30]. In order to have meaningful results in the prescriptions of
Einstein, Landau-Lifshitz and Papapetrou, it is necessary to
transform the metric in Cartesian coordinates. We transform this
metric in Cartesian coordinates by using
\begin{equation}
x=\rho\cos\phi,\quad y=\rho\sin\phi.
\end{equation}
The resulting metric in these coordinates will become
\begin{equation}
ds^2=e^{2\psi}dt^2-\frac{e^{2(\gamma-\psi)}}{\rho^2}(xdx+ydy)^2\nonumber\\
-\frac{e^{-2\psi}}{\rho^2}(xdy-ydx)^2-e^{2(\gamma-\psi)}dz^2.
\end{equation}

\section{Energy and Momentum in Einstein's Prescription}

The energy-momentum complex of Einstein [8] is given by
\begin{equation}
\Theta^b_a= \frac{1}{16 \pi}H^{bc}_{a,c},
\end{equation}
where
\begin{equation}
H^{bc}_a=\frac{g_{ad}}{\sqrt{-g}}[-g(g^{bd}g^{ce}
-g^{be}g^{cd})]_{,e},\quad a,b,c,d,e = 0,1,2,3.
\end{equation}
Here $\Theta^0_{0}$ is the energy density, $\Theta^i_{0}~
(i=1,2,3)$ are the momentum density components and $\Theta^0_{i}$
are the energy current density components. The Einstein
energy-momentum satisfies the local conservation laws
\begin{equation}
\frac{\partial \Theta^b_a}{\partial x^{b}}=0.
\end{equation}
The required non-vanishing components of $H^{bc}_{a}$ are
\begin{eqnarray}
H^{01}_{0}&=&\frac{x}{\rho^2}(e^{2\gamma}-1)
-\frac{2x}{\rho}(\gamma_\rho-2\psi_\rho),\\
H^{02}_{0}&=&\frac{y}{\rho^2}(e^{2\gamma}-1)
-\frac{2y}{\rho}(\gamma_\rho-2\psi_\rho),\\
H^{03}_{0}&=&2(\gamma_z-2\psi_z).
\end{eqnarray}
Using Eqs.(12)-(14) in Eq.(9), we obtain the energy and momentum
densities in Einstein's prescription
\begin{equation}
\Theta^0_{0}=\frac{1}{8\pi\rho}[\gamma_\rho(e^{2\gamma}
-1)-\rho\gamma_{\rho\rho}+2\psi_\rho
+2\rho\psi_{\rho\rho}+\rho\gamma_{zz}-2\rho\psi_{zz}],
\end{equation}
\begin{equation}
\Theta^i_0 =0.
\end{equation}
This gives momentum density zero and consequently momentum is
constant.
\section{Energy and Momentum in Landau-Lifshitz's Prescrition}

The Landau-Lifshitz [9] energy-momentum complex can be written as
\begin{equation}
L^{ab}= \frac{1}{16 \pi}\ell^{acbd}_{,cd},
\end{equation}
where
\begin{equation}
\ell^{acbd}= -g(g^{ab}g^{cd}-g^{ad}g^{cb}).
\end{equation}
$L^{ab}$ is symmetric with respect to its indices. $L^{00}$ is the
energy density and $L^{0i}$ are the momentum (energy current)
density components. $\ell^{abcd}$ has symmetries of the Riemann
curvature tensor. The local conservation laws for Landau-Lifshitz
energy-momentum complex turn out to be
\begin{equation}
\frac{\partial L^{ab}}{\partial x^{b}}=0.
\end{equation}
The required non-vanishing components of $\ell^{acbd}$ are
\begin{eqnarray}
\ell^{0101}&=&-\frac{1}{\rho^2}(y^2e^{4(\gamma-\psi)}
+x^2e^{2(\gamma-2\psi)}),\\
\ell^{0202}&=&-\frac{1}{\rho^2}(x^2e^{4(\gamma-\psi)}
+y^2e^{2(\gamma-2\psi)}),\\
\ell^{0102}&=&\frac{xy}{\rho^2}(e^{4(\gamma-\psi)}
-e^{2(\gamma-2\psi)}),\\
\ell^{0303}&=&-e^{2(\gamma-2\psi)}.
\end{eqnarray}
When we substitute these values in Eq.(17), it follows that the
energy density remains non-zero while momentum density components
vanish. These are given as follows
\begin{eqnarray}
L^{00}&=&\frac{-e^{2(\gamma-2\psi)}}{8\pi
\rho^2}[\{2(\gamma_\rho-\psi_\rho)\rho+1\}e^{2\gamma}-1
+\rho^2(\gamma_{\rho\rho}-2\psi_{\rho\rho}\nonumber\\
&+&\gamma_{zz} -2\psi_{zz})+2\rho^2\{(\gamma_\rho-2\psi_\rho)^2
+(\gamma_z-2\psi_z)^2\}],\nonumber\\
L^{0i}&=&0.
\end{eqnarray}

\section{Energy and Momentum in Papapetrou's Prescription}

We can write the prescription of Papapetrou [10] energy-momentum
distribution in the following way
\begin{equation}
\Omega^{ab}=\frac{1}{16\pi}N^{abcd}_{,cd},
\end{equation}
where
\begin{equation}
N^{abcd}=\sqrt{-g}(g^{ab}\eta^{cd}-g^{ac}\eta^{bd}
+g^{cd}\eta^{ab}-g^{bd}\eta^{ac}),
\end{equation}
and $\eta^{ab}$ is the Minkowski spacetime. It follows that the
energy-momentum complex satisfies the following local conservation
laws
\begin{equation}
\frac{\partial \Omega^{ab}}{\partial x^b}=0.
\end{equation}
$\Omega^{00}$ and $\Omega^{0i}$ represent the energy and momentum
(energy current) density components respectively. The required
non-vanishing components of $N^{abcd}$ are given by
\begin{eqnarray}
N^{0011}&=&-\frac{1}{\rho^2}[x^2+y^2e^{2\gamma}+\rho^2e^{2(\gamma-2\psi)}],\\
N^{0022}&=&-\frac{1}{\rho^2}[x^2e^{2\gamma}+y^2+\rho^2e^{2(\gamma-2\psi)}],\\
N^{0012}&=&\frac{xy}{\rho^2}(e^{2\gamma}-1),\\
N^{0033}&=&-1-e^{2(\gamma-2\psi)}.
\end{eqnarray}
Substituting Eqs.(28)-(31) in Eq.(25), we obtain the following
energy density and momentum density components
\begin{eqnarray}
\Omega^{00}&=&\frac{e^{2\gamma}}{8\pi\rho}[(1-e^{-4\psi})\gamma_\rho
+\{2\psi_\rho-\rho(\gamma_{\rho\rho}-2\psi_{\rho\rho}
+\gamma_{zz}-2\psi_{zz})\nonumber\\
&-&2\rho\{(\gamma_\rho-2\psi_\rho)^2+(\gamma_z-2\psi_z)^2\}\}e^{-4\psi}],\nonumber\\
\Omega^{0i}&=&0.
\end{eqnarray}

\section{Energy and Momentum in M\"{o}ller's Prescription}

The energy-momentum density components in M\"{o}ller's
prescription [14,15] are given as
\begin{equation}
M^b_a= \frac{1}{8\pi}K^{bc}_{a,c},
\end{equation}
where
\begin{equation}
K_a^{bc}= \sqrt{-g}(g_{ad,e}-g_{ae,d})g^{be}g^{cd}.
\end{equation}
Here $K^{bc}_{ a}$ is symmetric with respect to the indices.
$M^0_{0}$ is the energy density, $M^i_{0}$ are momentum density
components, and $M^0_{i}$ are the components of energy current
density. The M\"{o}ller energy-momentum satisfies the following
local conservation laws
\begin{equation}
\frac{\partial M^b_a}{\partial x^b}=0.
\end{equation}
Notice that M\"{o}ller's energy-momentum complex is independent of
coordinates. For the Weyl metric, we obtain the following
non-vanishing components of $K^{bc}_a$
\begin{equation}
K^{01}_0= 2\rho\psi_\rho,
\end{equation}
\begin{equation}
K^{03}_0= 2\rho\psi_z.
\end{equation}
When we make use of Eqs.(36) and (37) in Eq.(33), the energy and
momentum density components turn out to be
\begin{equation}
M^0_0=\frac{1}{4\pi}(\psi_\rho+\rho\psi_{\rho\rho}+\rho\psi_{zz}),
\end{equation}
\begin{equation}
M^i_0=0.
\end{equation}

\section{Discussion}

The debate on the localization of energy-momentum is an
interesting and a controversial problem. According to Misner et al
[31], energy can only be localized for spherical systems. In a
series of papers [32] Cooperstock et al. has presented a
hypothesis which says that, in a curved spacetime, energy and
momentum are confined to the regions of non-vanishing
energy-momentum tensor $T_a^b$ of the matter and all
non-gravitational fields. The results of Xulu [23,24] and the
recent results of Bringley [33] support this hypothesis. Also, in
the recent work, Virbhadra and his collaborators [18-22] have
shown that different energy-momentum complexes can provide
meaningful results. Keeping these points in mind, we have explored
the Weyl spacetime for the energy-momentum distribution.

In this paper, we are using prescriptions of Einstein,
Landau-Lifshitz, Papapetrou and M$\ddot{o}$ller to evaluate the
energy-momentum density components for the Weyl metric. From
Eqs.(14), (23), (32) and (38), it can be seen that the
energy-momentum densities are finite and well defined. We also
note that the energy density is different for the four different
prescriptions. However, momentum density components turn out to be
zero in all the prescriptions and consequently we obtain constant
momentum for this spacetime.

In recent papers [26-28], we have used Einstein and Papapetrou's
prescriptions to determine the energy-momentum distribution of
G$\ddot{o}$del and G$\ddot{o}$del type spacetimes. These results
do not coincide for the two different prescriptions. Here we have
found the spacetime in which the energy density is different for
the four prescriptions but the momentum become constant. We know
many examples for which the energy-momentum complexes can yield
the same results. It is mentioned here that these results turn out
to be the same [34] under the limiting case of the Curzon metric
which is a special solution of the Weyl metric. It would be
interesting to look for the reasons of this difference. It seems
that the basic problem of definition of energy-momentum in GR is
still there which needs to be resolved.

\newpage

{\bf \large References}

\begin{description}

\item{[1]} Lorentz, H.A.: Veral. Kon. Akad. Wet. Amsterdam {\bf 25}(1916)1380.

\item{[2]} Penrose, R.: {\it Proc. Roy. Soc.} London {\bf
A388}(1982)457;\\
{\it GR10 Conference}, eds. Bertotti, B., de Felice, F. and
Pascolini, A. Padova {\bf 1} (1983)607.

\item{[3]} Shaw, W.: {\it Proc. Roy. Soc.} London {\bf
A390}(1983)191;\\
in {\it Asymptotic Behaviour of Mass and Spacetime Geometry}, ed.
Flaherty, F.J. {\it Springer Lecture Notes in Physics 202}
(Berlin, 1984b);\\
Dray, T.: Class. Quant. Grav. {\bf L7}(1985)2;\\
Kelly, R.: Twistor Newsletter {\bf 20}(1985)11.

\item{[4]} Jeffryes, B.P.: {it GR11 Conference Absracts}, Jena (1986)539.

\item{[5]} Futamase, T. and Schutz, B.: Phys. Rev. {\bf D20}(1983)10.

\item{[6]} Bartnik, R.: {\it Proc. of the Fifth Marcel Grossmann Meeting
on General Relativity} ed. Blair, D.G. and Buckingham, M.J. (World
Scientific, 1989)399.

\item{[7]} Dirac, P.A.M.: {\it Proc. Roy. Soc.} London {\bf
A246}(1958)326;\\
Arnowitt, R. Deser, S. and Misner, C.W.: Phys. Rev. {\bf
117}(1960)1595.

\item{[8]} Trautman, A.: {\it Gravitation: An Introduction to Current
Research} ed. Witten, L. (Wiley, New York, 1962)169.

\item{[9]} Landau, L.D. and  Lifshitz, E.M.: {\it The Classical Theory of Fields}
(Addison-Wesley Press, 1962).

\item{[10]} Papapetrou, A.: {\it Proc. R. Irish Acad} {\bf A52}(1948)11.

\item{[11]} Bergman P.G: and Thompson R. Phys. Rev. {\bf89}(1958)400.

\item{[12]} Tolman R. C: Relativity, Thermodynamics and Cosmology,
(Oxford University Press, Oxford) p.{\bf 227}(1934).

\item{[13]} Weinberg, S.: {\it Gravitation and Cosmology} (Wiley, New York,
1972).

\item{[14]} M\"{o}ller, C.: Ann. Phys. (NY) {\bf 4}(1958)347.

\item{[15]} M\"{o}ller, C.: Ann. Phys. (NY) 12(1961)118.

\item{[16]} Komar, A. {\it Phys. Rev.} {\bf 113}(1959)934.

\item{[17]} Kovacs, D. {\it Gen. Relatv. and Grav.} {\bf 17}, 927
(1985); Novotny, J. {\it Gen. Relatv. and Grav.} {\bf 19}, 1043
(1987).

\item{[18]} Virbhadra, K.S.: Phys. Rev. {\bf D42}(1990)2919.

\item{[19]} Virbhadra, K.S.: Phys. Rev. {\bf D60}(1999)104041.

\item{[20]} Rosen, N. and Virbhadra, K.S.: Gen. Relati. Gravi. {\bf 25}(1993)429.

\item{[21]} Virbhadra, K.S. and Parikh, J.C.: Phys. Lett. {\bf B317}(1993)312.

\item{[22]} Virbhadra, K.S. and Parikh, J.C.: Phys. Lett. {\bf B331}(1994)302.

\item{[23]} Xulu, S.S.: Int. J. of Mod. Phys. {\bf A15}(2000)2979;
Mod. Phys. Lett. {\bf A15}(2000)1151 and reference therein.

\item{[24]} Xulu, S.S.: Astrophys. Space Sci. {\bf 283}(2003)23-32.

\item{[25]} Lessner, G. {\it Gen. Relativ. Grav.} {\bf 28}(1996)527.

\item{[26]} Sharif, M.: Int. J. of Mod. Phys. {\bf A18}(2003)4361;
Errata {\bf A19}(2004)1495.

\item{[27]} Sharif, M.: Int. J. of Mod. Phys. {\bf D13}(2004)1019;
Nuovo Cimento {\bf B119}(2004)463.

\item{[28]} Sharif, M. and Fatima, Tasnim: Int. J. of Mod. Phys. {\bf A}(2005).

\item{[29]} Weyl, H.: Ann. Phys. (Leipzig) {\bf 54}(1917)117;
{\bf 59}(1919)185;\\
Civita, Levi, L.: Atti. Acad. Naz. Lince Rend. Classe Sci. Fis.
Mat. e Nat., {\bf 28}(1919)101;\\
Synge, J.L.: {\it Relativity, the General Theory} (North-Holland
Pub. Co. Amsterdam, 1960);\\
Kramer, D., Stephani, H., MacCallum, M.A.H. and Hearlt, E.: {\it
Exact Solutions of Einstein's Field Equations} (Cambridge
University Press, 2003).

\item{[30]} Curzon, H.E.J.: {\it Proc. Math. Soc.} London {\bf 23}(1924)477.

\item{[31]} Misner,C.W., Thorne, K.S. and Wheeler, J.A. {\it
Gravitation} (W.H. Freeman, New York, 1973)603.

\item{[32]} Cooperstock, F.I. and Sarracino, R.S. {\it J. Phys. A.:
Math. Gen.} {\bf 11}(1978)877. \\
Cooperstock, F.I.: in {\it Topics
on Quantum Gravity and Beyond}, Essays in honour of Witten, L. on
his retirement, ed. Mansouri, F. and Scanio, J.J. (World
Scientific, Singapore, 1993); Mod. Phys. Lett. {\bf
A14}(1999)1531; Annals of Phys. {\bf
282}(2000)115;\\
Cooperstock, F.I. and Tieu, S.: Found. Phys. {\bf 33}(2003)1033.

\item{[33]} Bringley, T.: Mod. Phys. Lett. {\bf A17}(2002)157.

\item{[34]} Gad, R.M.: Mod. Phys. Lett. {\bf A19}(2004)1847.

\end{description}

\end{document}